\documentclass[11pt] {aa}

\usepackage{graphicx, subfig}
\usepackage{eqnarray,float}
\usepackage{multirow}
\usepackage{hhline}
\usepackage{siunitx}
\usepackage{natbib,twoopt, units}
\usepackage{lscape}
\usepackage[breaklinks=true]{hyperref} 
\bibpunct{(}{)}{;}{a}{}{,} 
\newcommandtwoopt{\citeads}[3][][]{\href{http://adsabs.harvard.edu/abs/}%
{\citealp[#1][#2]{#3}}} 
\newcommandtwoopt{\citepads}[3][][]{\href{http://adsabs.harvard.edu/abs/#3}%
{\citep[#1][#2]{#3}}} 
\newcommandtwoopt{\citetads}[3][][]{\href{http://adsabs.harvard.edu/abs/#3}%
{\citet[#1][#2]{#3}}} 
\newcommandtwoopt{\citeyearads}[3][][]%
{\href{http://adsabs.harvard.edu/abs/#3}{\citeyear[#1][#2]{#3}}}
\usepackage{hyperref}

\title{Gas composition of the main volatile elements in protoplanetary discs and its implication for planet formation}

\authorrunning{Thiabaud et al.}
\titlerunning{Gas composition of the main volatile elements in protoplanetary discs}

\author{Amaury~Thiabaud \inst{1,2}, Ulysse~Marboeuf\inst{1,2}, Yann Alibert \inst{1,2,3}, Ingo Leya\inst{1,2} \& Klaus Mezger \inst{1,4} }
\institute{$^1$Center for Space and Habitability, Universit\"{a}t Bern, CH-3012 Bern, Switzerland.\\   
Email: amaury.thiabaud@csh.unibe.ch\\
$^2$Physikalisches Institut, Universit\"{a}t Bern, CH-3012 Bern, Switzerland\\
$^3$Observatoire de Besan\c{c}on, 41 Avenue de l'Observatoire, 25000 Besan\c{c}on, France\\
$^4$Institut f\"{u}r Geologie, Universit\"{a}t Bern, CH-3012 Bern, Switzerland}

\date{Received ??; accepted 17/12/2014}

\begin{document}

	\abstract
   	{Direct observations of gaseous exoplanets reveal that their gas envelope has a higher  C/O ratio than that of the host star (e.g., Wasp 12-b). This has been explained by considering that the gas phase of the disc could be inhomogeneous, exceeding the stellar C/O ratio in regions where these planets formed; but few studies have considered the drift of the gas and planet migration.}
   	{We aim to derive the gas composition in planets through planet formation to evaluate if the formation of giant planets with an enriched C/O ratio is possible. The study focusses on the effects of different processes on the C/O ratio, such as the disc evolution, the drift of gas, and planet migration.}
	{We used our previous models for computing the chemical composition, together with a planet formation model, to which we added the composition and drift of the gas phase of the disc, which is composed of the main volatile species H$_2$O, CO, CO$_2$, NH$_3$, N$_2$, CH$_3$OH, CH$_4$, and H$_2$S, H$_2$ and He. The study focusses on the region where ice lines are present and influence the C/O ratio of the planets.}
   	{Modelling shows that the condensation of volatile species as a function of radial distance allows for C/O enrichment in specific parts of the protoplanetary disc of up to four times the solar value. This leads to the formation of planets that can be enriched in C/O in their envelope up to three times the solar value. Planet migration, gas phase evolution and disc irradiation enables the evolution of the initial C/O ratio that decreases in the outer part of the disc and increases in the inner part of the disc. The total C/O ratio of the planets is governed by the contribution of ices accreted, suggesting that high C/O ratios measured in planetary atmospheres are indicative of a lack of exchange of material between the core of a planet and its envelope or an observational bias. It also suggests that the observed C/O ratio is not representative of the total C/O ratio of the planet.}
   	 {}

   
   	\maketitle
	
	\section{Introduction}

		According to current understanding of stellar formation, the elemental ratios in a solar photosphere represent the elemental ratios of the nebula from which the central star formed (with very few exceptions like deuterium). Since planetesimals and planets are formed within this nebula, their composition is expected to match that of their host star. This assumption is probably true for highly refractory species, since their high condensation temperature places their condensation line close to the star \citep[][hereafter T14]{Lodders2003,Thiabaud2013}, but it may not apply to volatile species. Although the latter species will have no effect on the elemental ratios in the rocky part of planets (Fe/Si and Mg/Si), it is expected that they have an effect on the C/O ratio in the atmosphere as a result of differences in ice line positions during the early lifetime of the nebula while planets accrete \citep[][]{Oberg2011,Ali-Dib2014}. It is particularly difficult to estimate the amount of C and O in planetary bodies since they mostly occur in volatile species in a solar nebula \citep{Lodders2003}. Observations \citep{Morgan1980}, and simulations \citep[T14;][hereafter M14a and b; and references therein]{Marboeuf2014,Marboeuf2014a} show that the abundance of C in the bulk of the planets of the solar system is much lower than it was in the solar nebula \citep{Lodders2003}, requiring element fractionation processes during the early stages of planet formation. \\
		
		Similarly, recent astronomical observations of other stellar systems have shown that the C/O ratio in atmospheres of planets can deviate significantly from the host star value \citep[][]{Madhusudhan2012,Konopacky2013,Moses2013} and can also be different in the atmosphere of different planets in the same planetary system. \cite{Oberg2011} showed that these variations could be due to the different locations of condensation of H$_2$O, CO, and CO$_2$, resulting in an increase in the C/O ratios during condensation. This trend was subsequently reported by other studies \citep[see, e.g.,][]{Ali-Dib2014}. The main shortcoming of the studies of \cite{Oberg2011} and \cite{Ali-Dib2014} is that they did not take into account the planet migration during their formation or the gas phase evolution, which can both significantly change the C/O ratio in the atmosphere of growing planets. \\
		
		Several models have been published in an attempt to constrain the solid composition of planets. \cite{Bond2010,Bond2010a} and \cite{Carter-Bond2012} computed planet composition with different initial compositions, but their model did not take into account the formation pathway of planets in the protoplanetary discs and neglected the formation of volatile species (other than H$_2$O) in the solid phase. 
		\cite{Elser2012} were the first to attempt to include the composition of solids with a self-consistent planet formation model, but they were unable to reproduce some of the features of the solar system (high abundance of Fe of Mercury). Finally, M14a,b and T14 presented a model that includes the composition of both refractory and volatile species and considers several scenarios (energy contribution from stellar irradiation, formation of clathrates and organic compounds), in the full planet formation model of \cite{Alibert2013} (hereafter A13 - and references therein). However, these studies did not consider the accretion of gases other than He and H$_2$. The chemical composition of planets in their model was only derived from accretion of solid planetesimals (refractory and icy components). \\
		
		In the present study, we add to our previous study that modelled the solid composition of planetesimals and planets (M14a,b; T14) the possible accretion of an evolving gas disc composed of He, H$_2$, H$_2$O, CO, CO$_2$, CH$_3$OH, CH$_4$, NH$_3$, N$_2$, and H$_2$S. We focus the study on the critical region, that is, in the region of the different ice lines in a condensing nebula where a diversity of rocky planets can form. It is exactly this region where planets will form that will have vastly different evolution paths (e.g., core formation, atmosphere evolution, habitability). The aim is to study the implications of accreting such gas, including the possible increase in the C/O ratio, and to derive which physical processes can affect the composition of the formed planets. Section 2 briefly present the planet formation model of A13. Section 3 is dedicated to the chemical model used in this study. Results and discussions are presented in Sect. 4.
		
	\section{Planet formation model}
	
		The planet formation model is based on the work of A13,  which is based on the so-called core-accretion model of \cite{Pollack1996}. In this model, planets are formed in a protoplanetary disc. The planet core is formed by accretion of solid planetesimals until it eventually becomes massive enough to gravitationally bind some of the nebular gas, surrounding itself with a tenuous atmosphere. This formation model is able to produce different types of planets with both high (a few thousands of M$_{\oplus}$) and low masses (10$^{-2}$ M$_{\oplus}$).
		
		\subsection{Disc model}
		The protoplanetary disc model of \cite{Alibert2005d} is structured around three modules, that calculate the structure and evolution of a non-irradiated protoplanetary disc. The disc parameters are computed using a 1+1D model, which first resolves the vertical structure of the disc for each distance to the star. This calculation solves the hydrostatic equilibrium
		
		\begin{eqnarray}
			\frac{1}{\rho} \frac{\delta P}{\delta z}=-\Omega^{2} z,
		\end{eqnarray}
		
		\noindent the energy conservation
		
		\begin{eqnarray}
			\frac{\delta F}{\delta z}=\frac{9}{4} \rho \nu \Omega^{2},
		\end{eqnarray}
		
		\noindent and the diffusion equation for radiative flux
		
		\begin{eqnarray}
			F=\frac{-16\sigma T^3}{3 \kappa \rho} \frac{\delta T}{\delta z},
		\end{eqnarray}
		
		\noindent where $\rho$ is the volume density, $P$ the pressure, $\Omega$ the Keplerian velocity, $F$ the radiative flux, $\nu$ the viscosity, $\sigma$ the Stefan-Boltzmann constant, and $\kappa$ the opacity taken from \cite{Bell1994}. This calculation provides the vertically averaged viscosity as a function of the surface density in the disc \citep{Alibert2005d,Fortier2013}. This is then used to compute the radial structure and evolution as a function of this viscosity, the photo-evaporation, and the mass-accretion rate by solving the diffusion equation. \\
		
		The initial gas surface density (in g.cm$^{-2}$) is computed following
		
		\begin{eqnarray}
			\label{sigma_ini}	
				\Sigma=\Sigma_{0}\left(\frac{r}{a_{0}}\right)^{-\gamma}e^{(\frac{r}{a_{core}})^{2-\gamma}},
		\end{eqnarray}
		
		\noindent where a$_0$ is equal to 5.2 AU, a$_{\rm core}$ is the characteristic scaling radius (in AU), $\gamma$ is the power index, $r$ the distance to the star (in AU), and $\Sigma_0$=(2-$\gamma$)$\frac{M_{\rm disc}}{2\pi a_{\rm core}^{2-\gamma}a_0^\gamma}$ (in g.cm$^{-2}$). The values for a$_{\rm core}$, $\gamma$, and $\Sigma_0$ are varied for every disc, and are derived from the observations of \cite{Andrews2010}. The mass of the gas disc is inferred from the observations of solid discs through a dust-to-gas ratio randomly chosen to lie between 0.008 and 0.1 from a list of $\sim$1000 CORALIE targets (see A13 for more details) to produce 500 different discs. The characteristics of the disc models are presented in Table \ref{charadisc}, assuming that the gas-to-dust ratio is 100:1 for M$_{\rm disc}$.\\
		
		The parametric surface density profile (Eq. \ref{sigma_ini}) is only used as an initial condition (t=0). This initial profile evolves with time and the surface density profile at t!= 0 is recomputed at each timestep, following
		\begin{eqnarray}
			\label{sigma_evol}
			\frac{\delta \Sigma}{\delta t}=\frac{3}{r}\frac{\delta}{\delta r} \left(r^{\rm 1/2} \frac{\delta}{\delta r}(\nu \Sigma r^{\rm 1/2})\right)+Q_{\rm acc}+Q_{\rm ph},
		\end{eqnarray}
		
		\noindent where $Q_{\rm acc}$ is an additional term related to the accretion by planets and $Q_{\rm ph}$ an additional term related to photo-evaporation.
		
		Solving the equations and knowing the surface density at distance $r$ constrains the temperature and pressure at each distance and time of the disc. 
		
\begin{figure}
				\includegraphics[width=\columnwidth]{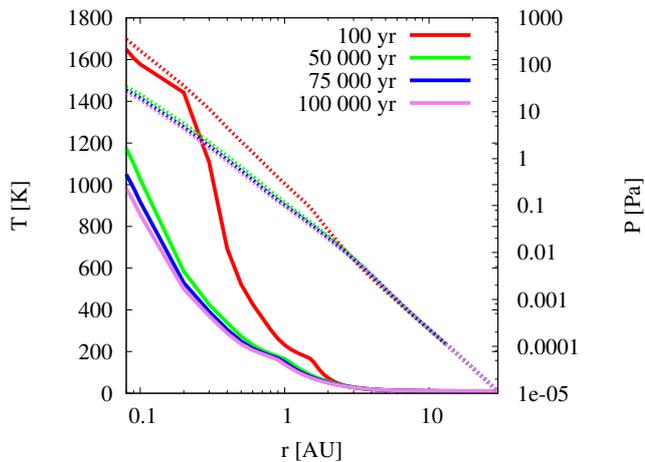}
				\caption{\label{rtp_evol} Time evolution of the midplane temperature (solid lines) and pressure (dotted lines) of disc \#1 ($\Sigma_0$=95.8 g.cm$^{-2}$, a$_{\rm core}$=46 AU, $\gamma$=0.9) in a case without irradiation.}
			\end{figure}

		 \begin{table}[ht]
				\centering
				\caption{\label{charadisc} Characteristics of disc models from A13, assuming for M$_{\rm disc}$ that the gas to dust ratio is 100:1.}
				\begin{tabular}{|c|c|c|c|}
					\hline
					Disc & M$_{\rm disc}$ (M$_{\odot})$ & a$_{\rm core}$ (AU) & $\gamma$ \\
					\hline
					1 & 0.029 & 46 & 0.9 \\
					2 & 0.117 & 127 & 0.9 \\
					3 & 0.143 & 198 & 0.7 \\
					4 & 0.028 & 126 & 0.4 \\
					5 & 0.136 & 80 & 0.9 \\
					6 & 0.077 & 153 & 1.0 \\
					7 & 0.029 & 33 & 0.8 \\
					8 & 0.004 & 20 & 0.8 \\
					9 & 0.012 & 26 & 1.0 \\
					10 & 0.007 & 26 & 1.1 \\
					11 & 0.007 & 38 & 1.1 \\
					12 & 0.011 & 14 & 0.8 \\
					\hline
				\end{tabular}
			\end{table} 
			 Figure \ref{rtp_evol} shows how the temperatures and pressures at the midplane of a particular disc ($\Sigma_0$=95.8 g.cm$^{-2}$, a$_{\rm core}$=46 AU, $\gamma$=0.9) evolves through time.
			 
			\paragraph{\textbf{Irradiation}} The planet formation and their final composition is influenced by irradiation. The  model of planet population synthesis of \cite{Alibert2005d} and \cite{Mordasini2009,Mordasini2009a} was used in \cite{Fouchet2012} to reveal the effects of irradiation on planet formation. The disc model is similar to that of A13 (1+1D viscous disc), but the irradiation is included by modifying the temperature boundary condition at the disc surface. This temperature is set to be $T^4_{\rm s}=T^4_{\rm s, no irr}+T^4_{\rm s, irr}$, where the irradiation temperature (in K) is given by \citep[][]{Hueso2005}
		
		\begin{eqnarray}	
			T_{\rm s, irr}=T_{\ast} \left[\frac{2}{3\pi}\left(\!\!\frac{R_{\ast}}{r}\!\!\right)^{\!\!3}\!\!+\frac{1}{2}\left(\!\!\frac{R_{\ast}}{r}\!\!\right)^{\!\!2}\!\!\left(\!\!\frac{H_{\rm P}}{r}\!\!\right)\left(\!\!\frac{d ln H_{\rm P}}{d ln r}-1\!\!\right)\right]^{1/4},
		\end{eqnarray}
		
		where $R_{\ast}$ and $T_{\ast}$ are the young stellar radius ($R_{\ast}$ = 1.75 $R_{\odot})$ and temperature ($T_{\ast}$ = 5000 K), and $H_{\rm P}$ is the pressure scale height. In this work, both models with and without irradiation from the central star are run.

		\subsection{Gas phase evolution} \label{Drift}
		In contrast to planetesimals, it cannot be assumed that the gas of the disc does not drift. The gas evolves with time with most of it accreting onto the star, at least for close-in regions.
		Because of the viscosity acting in the disc, two sheets of gas will interact through friction along their interface as a result of differential rotation. Since the gas is orbiting roughly in a Keplerian movement, the inner sheet rotates faster than the outer sheet, resulting in an exchange of angular momentum to decrease the speed of the inner sheet while increasing the speed of the outer one. It thus simultaneously allows gas to flow inwards. This phenomenon has been described by the theory of accretion discs by \cite{Lynden-Bell1974}. They derived the radial velocity of the gas $v_{\rm drift}$ (in cm.s$^{-1}$) due to mass and angular momentum conservation in the presence of a viscosity $\nu$ to be
		
		\begin{eqnarray} \label{equa_drift}
			v_{\rm drift} = - \frac{3}{\Sigma r^{\rm 1/2}} \frac{\delta}{\delta r} (\nu \Sigma r^{\rm 1/2}).
		\end{eqnarray}
		
\begin{figure}
			\includegraphics[width=\columnwidth]{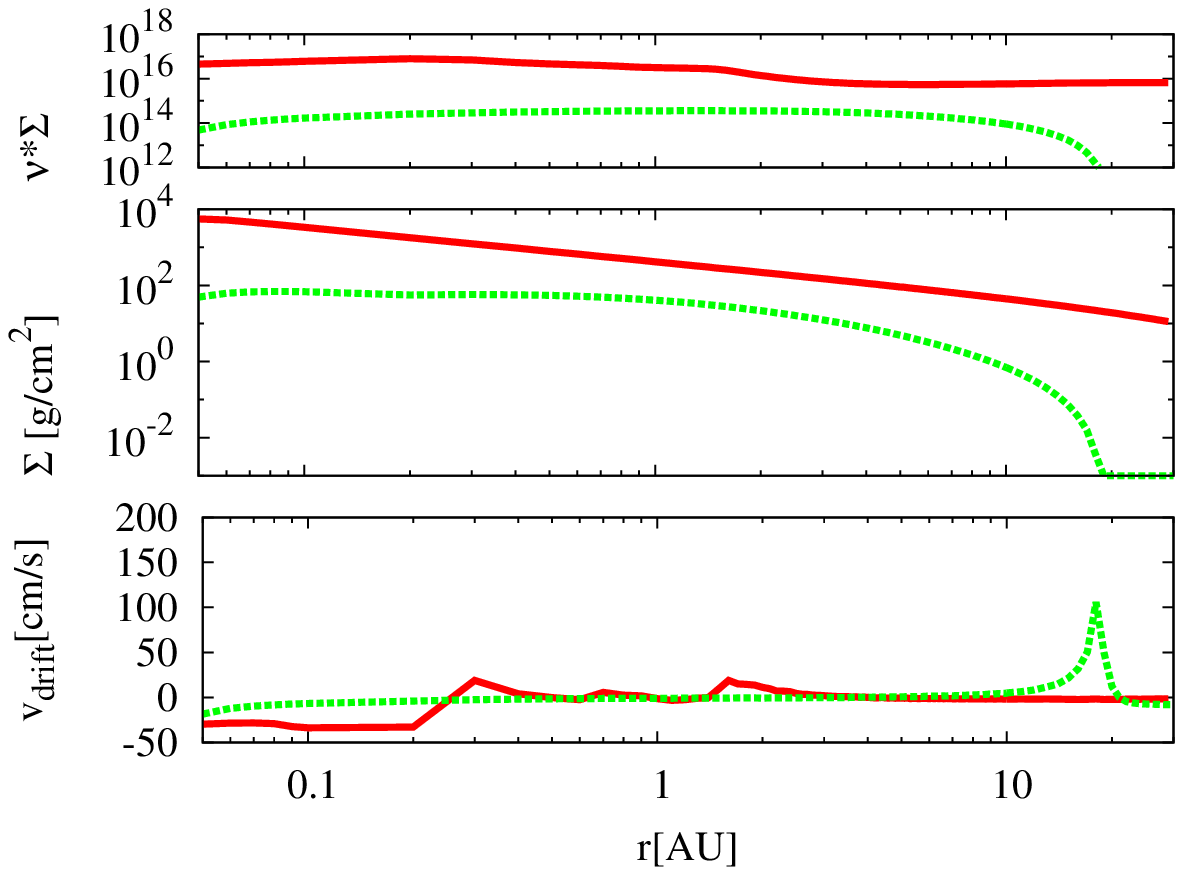} \\
			\includegraphics[width=\columnwidth]{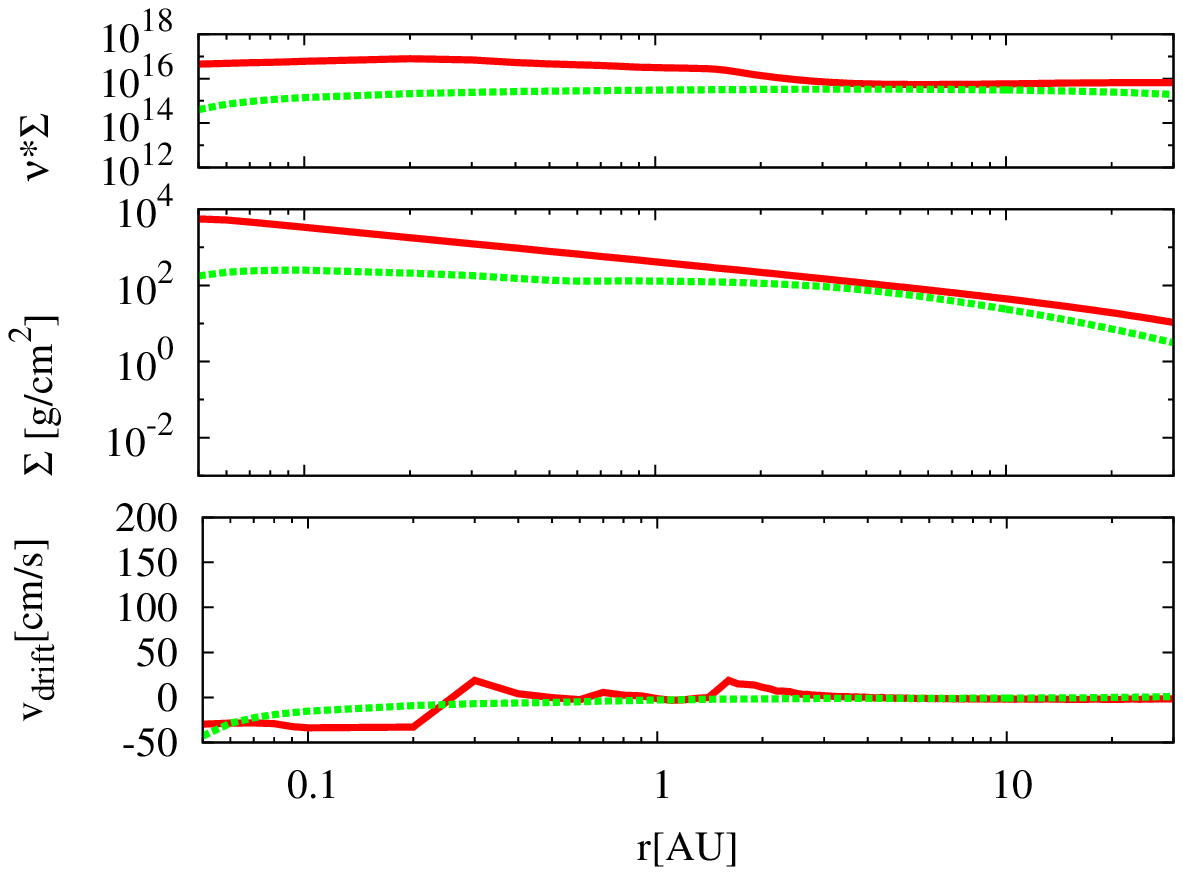}
			\caption{\label{v_drift} Radial profiles of the drift velocity of the gas, the surface density $\Sigma$ of the gas, and $\nu \Sigma$ for the model without irradiation (top) and with irradiation (bottom). The plots are shown for disc \#1 ($\Sigma_0$=95.8 g.cm$^{-2}$, a$_{\rm core}$=46 AU, $\gamma$=0.9) at the beginning (red line) and the end (green line) of the simulations (t=10Myr). The peak of $v_{\rm drift}$ at $\sim$ 20 AU for the model without irradiation arises because $\Sigma$ has reached its lowest value.}
		\end{figure}

		The idea of the present study is to take the initial profile of the gas abundances of H$_2$O, CO, CO$_2$, CH$_3$OH, CH$_4$, NH$_3$, N$_2$, and H$_2$S and evolve this profile through time via the drift velocity. Since $\Sigma$ changes with time (see Eq. \ref{sigma_evol}), this velocity is variable for a specific radius, as shown in Fig. \ref{v_drift}. The diffusion is hereby neglected in contrast to  \cite{Ali-Dib2014}. However, the disc structure and planet formation is computed as a function of time, whereas \cite{Ali-Dib2014} have a more realistic diffusion model, but only computed snapshots of the disc in time.
		
		The cumulative distribution of disc lifetimes is assumed to decay exponetially with a characteristic time of 2.5 Myr \citep[][A13]{Fortier2013}. The photoevaporation rate is then adjusted so that the protoplanetary disc mass reaches 10$^{-5}$M$_\odot$ at the selected disc lifetime, when calculations are stopped. The longest lifetime of a disc is 10 Myr (see A13 for more details).

		\subsection{Planet formation}
		The protoplanetary disc is assumed to be composed of gas with planetesimals and ten planetary embryos of lunar mass uniformly distributed in log between 0.04 and 30 AU, the region of planet formation. Note that the planetesimals are assumed to be already formed in the protoplanetary discs as their formation process is currently unknown, and that they are also assumed to be large enough (roughly 1 km) so that they do not drift. \\
		The model includes several processes: 
			\begin{itemize}
				\item Accretion of planetesimals, growing the solid core of the planet \citep[][]{Alibert2005d,Fortier2013}.
				\item Accretion of gas. If the core of the planet reaches the critical core mass, the gas is accreted in a runaway fashion, leading to the formation of gas giants \citep[][]{Alibert2005d}.
				\item Planet-disc interactions, which contributes to migration (type I or type II) of planets \citep[][]{Alibert2005d,Mordasini2009a,Mordasini2009}.
				\item Planet-planetesimal interactions, by which the planetesimal is excited by the planet.
				\item gravitational planet-planet interactions, making it possible to catch a pair of planets (or more) in mean motion resonance, as well as pushing each other. 
			\end{itemize}

	\section{Chemical model}
		\subsection{Solid phase}
		We follow M14a and T14  and assumed that the solid composition is set by the initial conditions ($\Sigma$, T, P) of the disc given by the planet formation model of A13. For both refractory and volatile materials, the condensation sequence is computed to derive the chemical composition of planetesimals. This condensation sequence is stopped at the temperature of the disc at the corresponding distance to the star.
		
		M14a,b and T14 have run several models with different compositions, taking into account the formation of several refractory species (a list of these species is provided in Table 1 of T14) and eight main volatile molecules: H$_2$O, CO, CO$_2$, CH$_3$OH, CH$_4$, NH$_3$, N$_2$, and H$_2$S - see Table \ref{ices_prop}. These species are the most abundant volatile molecules observed in the interstellar medium (ISM) \citep[][and references in M14]{Gibb2000,Gibb2000a,Gibb2004a,VanDishoeck2004,Whittet2007,Boogert2011,Mumma2011}, and in solar cometary comas \citep{Bockelee-Morvan2004,Crovisier2006,Mumma2011}. Previous studies (T14, M14a,b) also explored the possible formation of clathrates and refractory organic compounds and studied the effect of irradiation. T14 showed that a model without refractory organic compounds better reproduces the observations made of the solar system. The formation of clathrates in protoplanetary discs is still uncertain. For these reasons, we did not include the formation scenarios of clathrates and refractory organic compounds.
						
		\begin{table}[ht]
			\centering
			\begin{tabular}{|c|c|c|c|}
				\hline
				Specie & T$_{\rm cond}$ (K) &  \multicolumn{2}{|c|}{Nebulae abundance (mass/H$_2$)} \\
				\hline
				 & & CO:H$_2$O=0.2 & CO:H$_2$O=1 \\
				\hline
				H$_2$O & 152-173 & 8.16 10$^{-3}$ & 5.62 10$^{-3}$ \\
				CH$_3$OH & 127-143 & 2.17 10$^{-3}$ & 1.5 10$^{-3}$ \\
				NH$_3$ & 88-99 & 5.39 10$^{-4}$ & 3.72 10$^{-4}$ \\
				CO$_2$ & 77-86 & 3.99 10$^{-3}$ & 2.75 10$^{-3}$ \\
				H$_2$S & 67-75 & 3.09 10$^{-4}$ & 2.13 10$^{-4}$ \\
				CH$_4$ & 31-60 & 4.35 10$^{-4}$ & 3.00 10$^{-4}$ \\
				CO & 28-60 & 2.54 10$^{-3}$ & 8.75 10$^{-3}$ \\
				N$_2$ & 23-60 & 8.89 10$^{-4}$ & 6.12 10$^{-4}$ \\ 
				\hline
			\end{tabular}
			\caption{\label{ices_prop} Condensation temperatures and nebulae abundances of volatile species in the model.}
		\end{table}

		Irradiation, however, has an impact on the results as it adds another heating process. The ice lines were thus shifted outwards and the planets formed were more rocky than in the model without irradiation. It is consequently obvious that irradiation also plays a role in the gas composition of both discs and planets. Two cases were therefore considered, one with and one without irradiation.
		
		Two models shown in M14a,b were also computed in this study. Since the two most abundant C-bearing volatile compounds CO and CO$_2$ have a CO:CO$_2$ ice ratio ranging from about 1 to 4 in all types of sources \citep[][]{Gibb2004}, a CO:CO$_2$ molar ratio of 1:1 and 5:1 was adopted in this study as lowest and highest values. These values are consistent with the ISM measurements taking into account the contributions of both gas and ice phases \citep[][]{Mumma2011}. This leads to CO:H$_2$O molar ratios of 0.2 (resp. CO:CO$_2$=1:1) and 1 (resp. CO:CO$_2$=5:1).
		
		More details of the computation of refractory species are given in T14, and volatile species are computed in M14a,b.
		
		\subsection{Gas phase}
		Owing to their high condensation temperatures, refractory species in the gas phase are located close to the star (inside the ice line, a few AU). By the time the planet is massive enough to bind some gas, most of the gas within this location has been accreted onto the star. It is thus assumed in this contribution that the enrichment of refractory species due to gas accretion is negligible. Therefore we only included volatile molecules that condense at larger distance from the star\footnote{Note that the refractory enrichment can also come from grains closely coupled with the gas. This contribution is expected to be lower than solar however.}. 
		Volatile species in the gas phase are determined by local pressure and temperature and by partial pressure of the molecule. If the equilibrium pressure of the species is lower than its partial pressure in the gas, the species are assumed to condense (see M14a,b), otherwise they are left in the gas phase.  If species remain in the gas phase, their composition is given by
		
		 \begin{eqnarray}
		 	N_{\rm X} = Y_{\rm X} \  .\ N_{\rm H_2},
		 \end{eqnarray}
		 
		 where $N_{\rm X}$ is the abundance in the gas of species X, $Y_{\rm X}$ is the ISM abundance of species X relative to H$_2$, and $N_{\rm H_2}$ is the abundance of $H_2$.

	\section{Results}
		\subsection{Gas phase evolution}
\begin{figure}
				\includegraphics[width=\columnwidth]{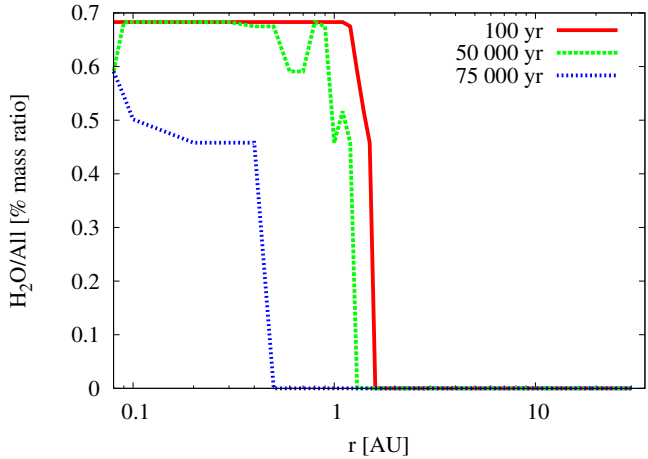}
				\caption{\label{water_evo} Time evolution of the gas phase abundance of H$_2$O relative to all gas for disc \#1 ($\Sigma_0$=95.8 g.cm$^{-2}$, a$_{\rm core}$=46 AU, $\gamma$=0.9). The plot is shown for CO/H$_2$O = 0.2 and without irradiation. Note that the 100 000 yr curve is set to zero since H$_2$O has disappeared from the disc by this time.}
			\end{figure}

\begin{figure}
				\includegraphics[width=\columnwidth]{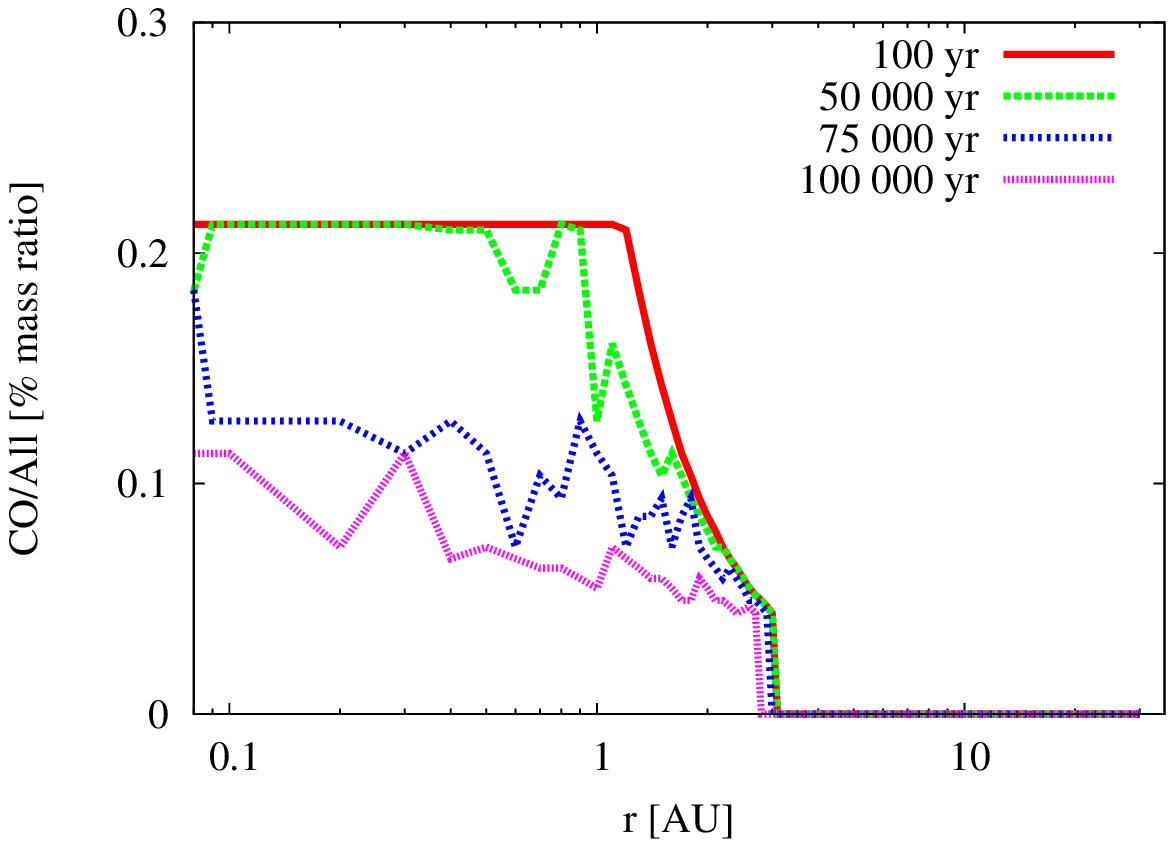}
				\caption{\label{CO_evo} Time evolution of the gas phase abundance of CO relative to all gas for disc \#1 (analogous to Fig. \ref{water_evo}).}
			\end{figure}

\begin{figure}
				\includegraphics[width=\columnwidth]{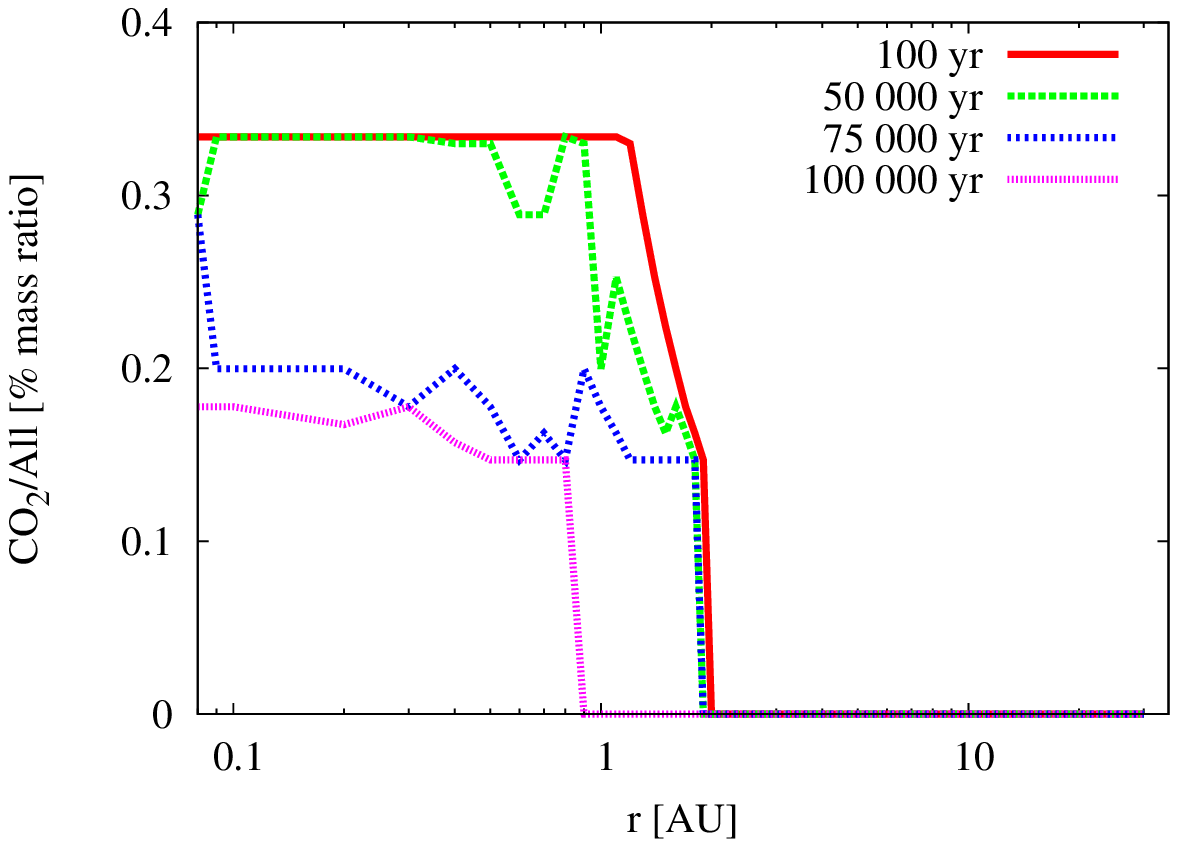}
				\caption{\label{CO2_evo} Time evolution of the gas phase abundance of CO$_2$ relative to all gas for disc \#1 (analogous to Fig. \ref{water_evo}).}
			\end{figure}

\begin{figure}
				\includegraphics[width=\columnwidth]{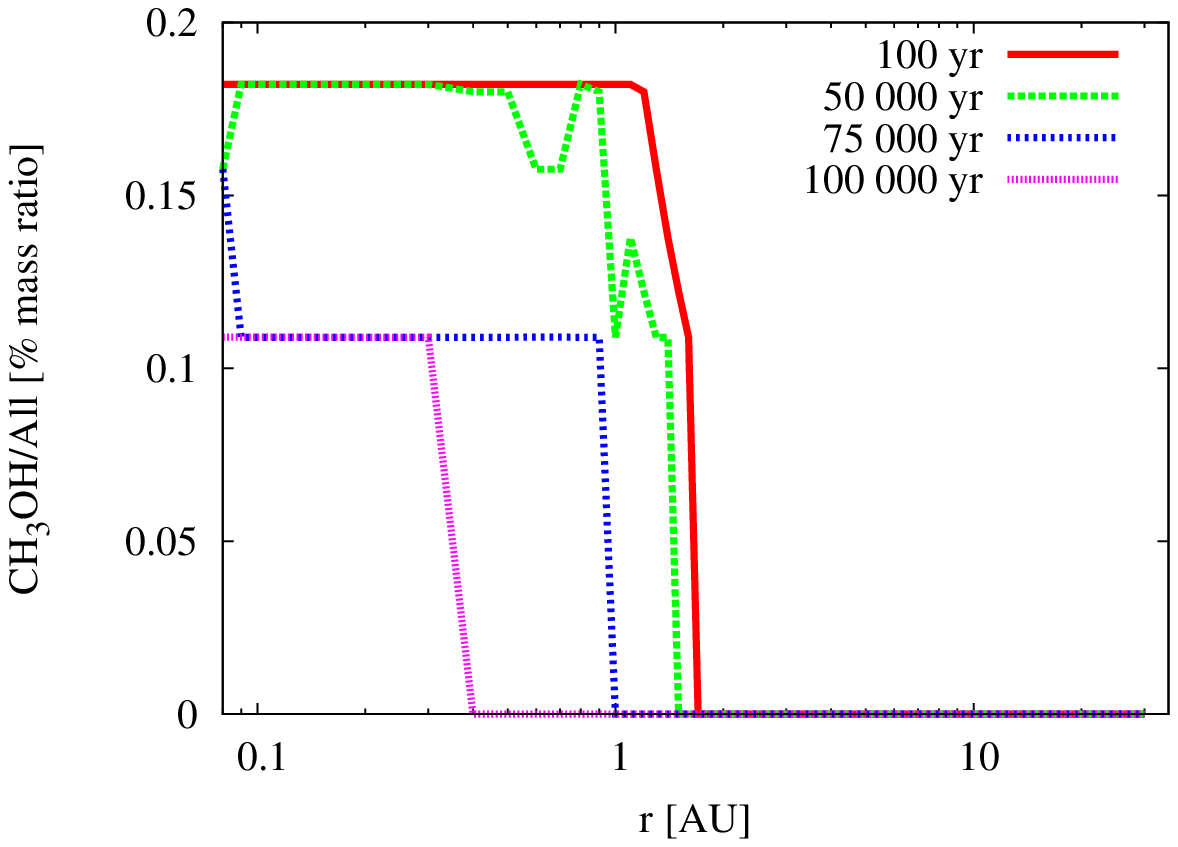}
				\caption{\label{CH3OH_evo} Time evolution of the gas phase abundance of CH$_3$OH relative to all gas for disc \#1 (analogous to Fig. \ref{water_evo}).}
			\end{figure}

\begin{figure}
				\includegraphics[width=\columnwidth]{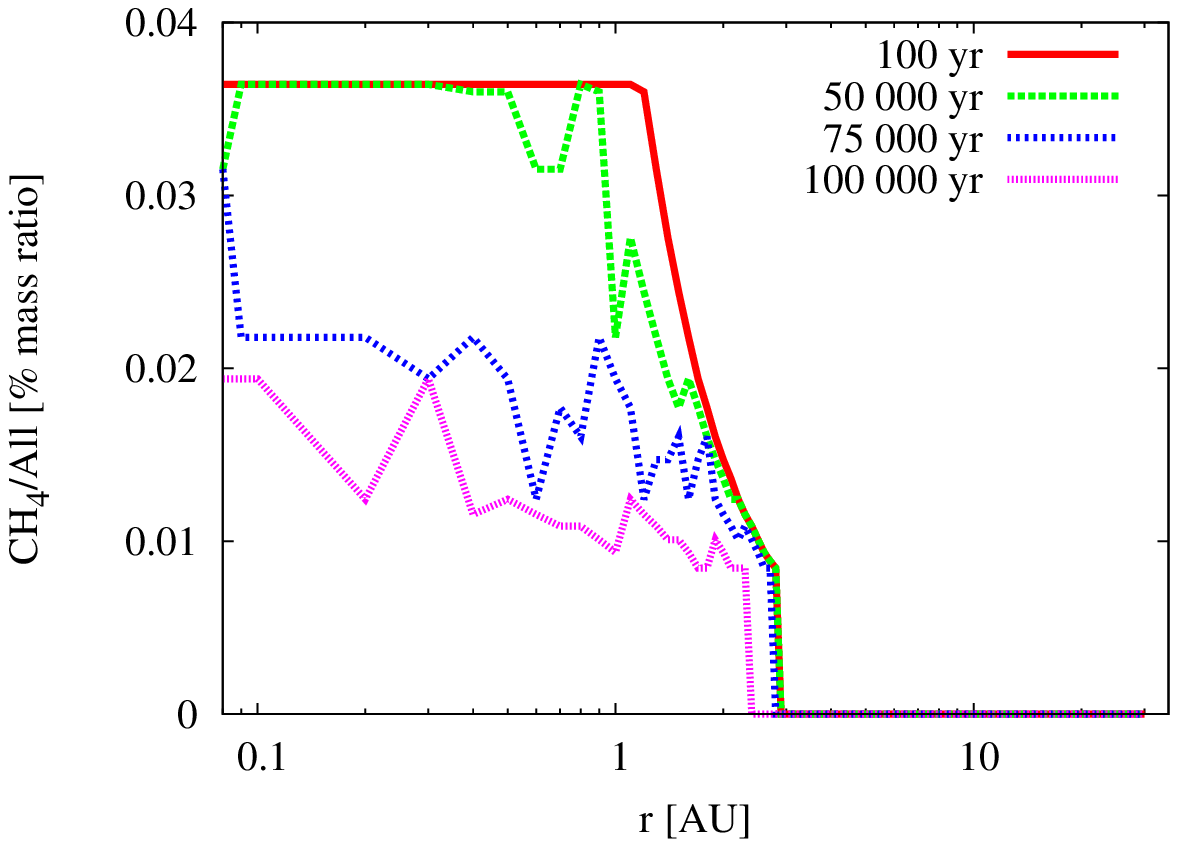}
				\caption{\label{CH4_evo} Time evolution of the gas phase abundance of CH$_4$ relative to all gas for disc \#1 (analogous to Fig. \ref{water_evo}).}
			\end{figure}

			As stated in Sect. \ref{Drift}, the initial profile of gas abundances of the eight main volatile molecules is computed and then evolves with the drift velocity given by Eq. \ref{equa_drift}. Figures \ref{water_evo} to \ref{CH4_evo} show the evolution of the gas abundances of these species at four different times (100, 50 000, 75 000, and 100 000 yr) for disc \#1 of A13 ($\Sigma_0$=95.8 g.cm$^{-2}$, a$_{\rm core}$=46 AU, $\gamma$=0.9) with CO/H$_2$O = 0.2, and without irradiation. The most abundant molecule in the gas phase is H$_2$O between the water ice line and the star, followed by CO$_2$, CO, and CH$_3$OH. Water in the gas phase is the only compound to disappear within 100 000 yr from this disc. The H$_2$O ice line being the closest to the star, the time needed to drift the gas from the H$_2$O ice line is shorter than the time needed to drift the other gases from their respective ice lines.
			
			The plots are similar in the model with irradiation, the drop in gaseous abundances occurs at a larger distance from the star, which induces a loss of the water later in time (about 10 000 - 20 000 yr later), which is also seen in discs with high masses in the model without irradiation. 
				
			\subsection{C/O gas enrichment}
\begin{figure}
				\includegraphics[width=\columnwidth]{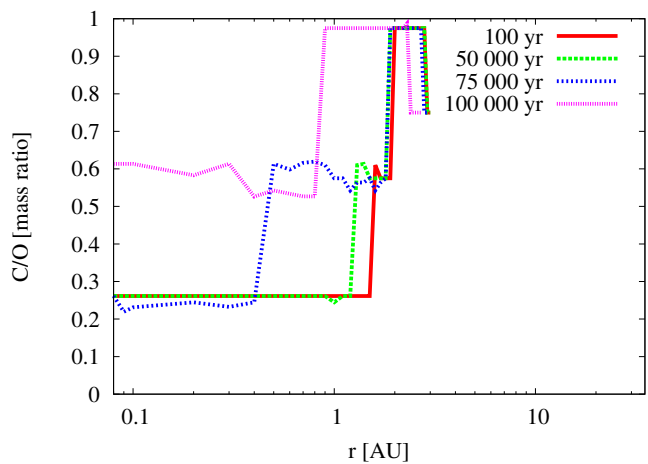}
				\caption{\label{CO_ratio_20_nirr} C/O mass ratio in the gas phase for the disc \#1 of A13 ($\Sigma_0$=95.8 g.cm$^{-2}$, a$_{\rm core}$=46 AU, $\gamma$=0.9), plotted after 100, 50 000, 75 000, and 100 000 years of evolution. The plot is shown for CO/H$_2$O = 0.2 and without irradiation.}
			\end{figure}

			The disappearance of different species in the gas phase through time is a phenomenon that can lead to an increase or decrease of the C/O ratio in the gas phase. This creates a dichotomy in time in the C/O ratio in the discs and thus in the atmosphere of planets in the same planetary system. A spatial dichotomy is also observable due to the location of embryo formation. Figure \ref{CO_ratio_20_nirr} shows the C/O ratio in the gas phase of disc \#1 of A13 ($\Sigma_0$=95.8 g.cm$^{-2}$, a$_{\rm core}$=46 AU, $\gamma$=0.9), plotted for the same time steps as Figs. \ref{water_evo} to \ref{CH4_evo}. 
			
			The first time step is representative of the initial conditions in the gas disc. Until water condenses at roughly 1.5 AU, the C/O ratio stays constant. The condensation of water induces a depletion in O in the gas phase, which increases the C/O ratio by roughly a factor 2. The condensation of methanol, whose ice line is located $\sim$0.1 AU farther out, induces a weaker decrease of the C/O ratio. The next molecule that condenses is NH$_3$ whose condensation does not affect the C/O ratio, in contrast to the condensation of CO$_2$ at 2 AU. The depletion of O is thus greater than the depletion in C, resulting in a new increase in the C/O ratio. Finally, the two last decreases are due to the condensation of CH$_4$ around 2.8 AU and of CO at 3.1 AU. 
		
			The C/O ratio increases after 100 000 yr close to the star. This is due to the disappearance of  water in the gas phase after 100 000 yr (see Figs. \ref{water_evo} and \ref{CO_ratio_20_nirr}). Consequently, if a planet has been accreting most of its gas after 100 000 yr, an increase compared to other planets could be observed. This is another process that can account for differences in the C/O ratios in  planetary atmospheres.
			Previous studies did not include CH$_4$ in the gas phase of the disc. Its presence is very likely, however, given the composition of planetary and cometary bodies, and has been detected in the disc of GV Tau N \citep{Gibb2013}. The inclusion of CH$_4$ in the models indicates an increase in the C/O ratio that has not been observed in previous studies \citep[][]{Oberg2011,Ali-Dib2014}. This increase implies an enrichment of the C/O ratio by a factor of 3 to 4 compared to the solar value at distances where CH$_4$ and CO are the only C- and O-bearing species in the gas phase.
		
		\subsection{Gas composition of planets}
			\subsubsection{Contribution of the gas phase}
\begin{figure*} 
				\includegraphics[width=2\columnwidth]{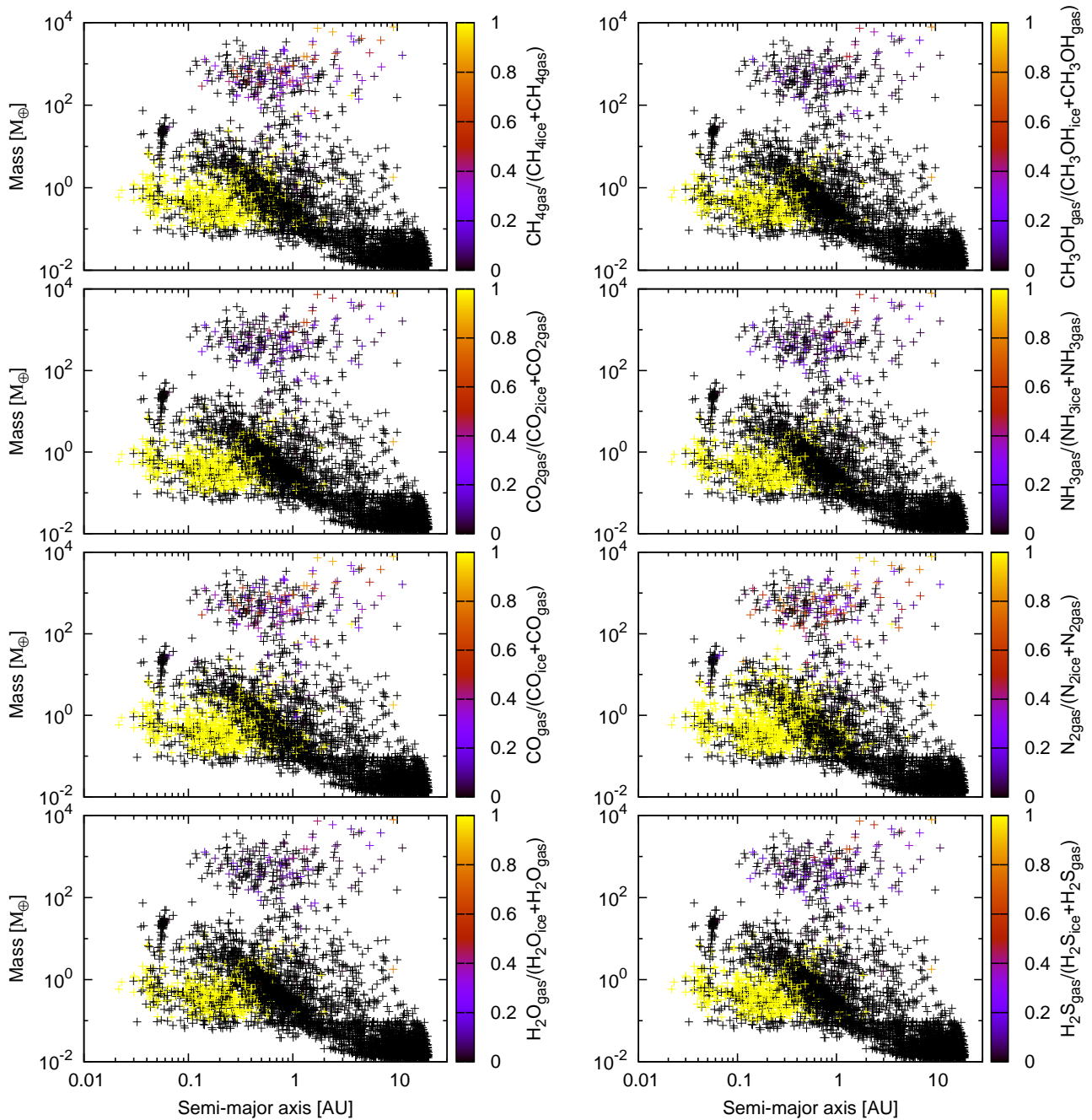}
				\caption{\label{proportions} Gas phase fraction of each species related to the total amount (gas+solid) of this species in an a-M diagram. Plots are shown for CO/H$_2$O = 0.2 without irradiation and are identical for CO/H$_2$O = 1. They have similar features in an model with irradiation.}
			\end{figure*}

			The amount of volatile species incorporated into planets depends on both the amount of gas and of solids accreted by the planet. Figure \ref{proportions} shows the amount of gaseous volatile species related to the total amount of volatile species (gas and solid) in the planet. \\
				
			For planets with a mass lower than 100 M$_{\oplus}$ the contribution of the gas phase is negligible ($<$ 1\%) in planets where ices are present (planets located at distances beyond 0.5 AU), meaning that the total abundances of the planet will be determined by its solid composition. However, for rocky planets (see Fig. 3 of T14) that did not accrete ices, the total amount of volatile species is defined by the low amounts of accreted gas as expected from the terrestrial planets of the solar system, whose volatile abundance in solids is lower than 0.1\% \citep[][]{Morgan1980}. \\
			
			 For 50\% of the giant planets ( $>$ 100 M$_{\oplus}$), the abundances of volatile species in the planet is dictated by the accreted amount of ices. This is because they either formed outside  the N$_2$ ice line (the last species to condense) in a low-mass disc or they accreted at a time when only H$_2$ and He were left in the gas phase of the disc. 
			 
			 For the other 50\%, the amount of volatile compounds accreted by the planet in the gas phase contributes between 20\% and 100\% of the total amount of volatile species in the planet (gas+ices).This is because gas giants are more easily formed in massive discs. Hence there is an increase in the temperature profiles relative to less massive discs, an increase in the abundances of the gas phase, and a decrease of the ice abundances. Such giant planets have a more highly refractory core than other giant planets, thus increasing the contribution of the gas part to the total abundance of volatile species. The time of accretion plays a role as well, that is, if the planet accretes gas just before and after the disappearance of the species in the gas phase of the disc.
			 
			  Physical parameters of the disc (surface density, temperature, mass, etc.) in which planets are formed and the time of accretion of gas are thus important for determining the relative abundance of species in gas and solid phases.

			\subsubsection{Contributions to the C/O ratio}
\begin{figure*}		
				\includegraphics[width=2\columnwidth]{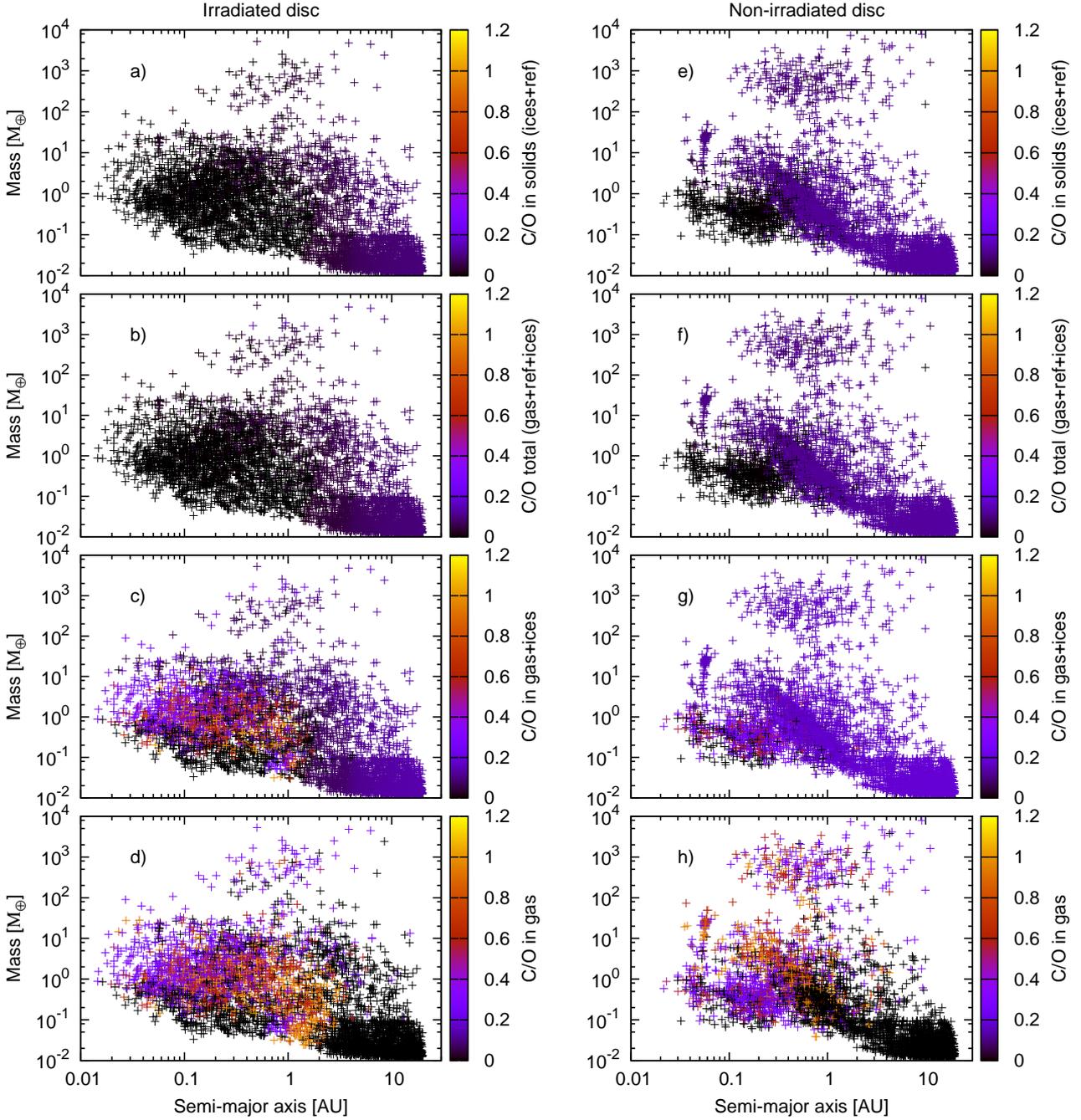}
				\caption{\label{CO_20_plan} C/O mass ratio in planets of the planet population synthesis of A13 for the model with irradiation (left) and without irradiation (right) in a mass-semimajor axis diagram. Different contribution to the C/O ratio in planets are shown. a-e) the C/O ratio in solids (ices and minerals); b-f) the total C/O ratio (in solids and in gas phase); c-g) the C/O ratio in atmospheres if ices are assumed to fully desorb into the gas phase; and d-h) the C/O ratio obtained with the gas phase of the protoplanetary disc alone. The plot is shown for CO/H$_2$O = 0.2.}
			\end{figure*}

			Figure \ref{CO_20_plan} shows the C/O ratio in planets formed in irradiated and non-irradiated discs for CO/H$_2$O = 0.2 \footnote{The plots for CO/H$_2$O = 1 are not shown here, but note that the CO/H$_2$O = 1 scenario gives a slight increase of $\sim$0.05 of the C/O ratio in rocky and giant planets whose C/O is around 0.4 in the gas phase in Fig. \ref{CO_20_plan}; and a slight decrease to 0.8 for rocky and giant planets with a C/O ratio higher than 0.9 in Fig. \ref{CO_20_plan}. The conclusions are unchanged in this case.}.
	
			The gas accreted by planets can increase the atmospheric C/O ratio by as much as three to four times the solar value of 0.37 by mass \citep[][]{Lodders2003}, showing that it is possible to form planets with a very different C/O ratio in their atmosphere compared to the host star, as observed in WASP-12b \citep[see][and reference therein]{Moses2013}. However, the contribution of gas to the total C/O ratio (considering C and O atoms in ices, minerals, and gas) is in most cases negligible (Fig. \ref{CO_20_plan} b-f), meaning that only the solid components are constraining the total C/O ratio in planets. This suggests that the observed C/O ratio in atmospheres may not be representative of the total C/O ratio of the planet.
			
			If the ices are assumed to sublimate into the atmosphere (Fig. \ref{CO_20_plan} c and g), the C/O ratio in the envelope is largely dominated by their contribution for the model without irradiation, leaving only but a few differences in the C/O ratio between planets (most planets have a ratio around 0.2). The contribution of ices is decreased if the disc is irradiated (apart from far planets whose location is $>$ 5 AU) because the abundance of ices in planets formed in such discs is much lower (see T14, M14a,b). In the region between 1 and 2 AU ices and accreted gas coexist and the contribution of the ices dominates. 
			This suggests that the high C/O ratios that are observed in hot Jupiters could come from a lack of exchange of material between the planet core and its envelope, or it might highlight an observational bias towards layered-type atmospheres. If ices sublimate, the C/O ratio will approach the stellar value, but if the volatile species in the envelope condense, the C/O ratio approaches 0. 
			This second scenario (lack of exchange and/or layered-type atmosphere) is consistent with the observed C/O ratios in hot Jupiters. However, planetesimals probably suffer ablation in the atmosphere of such planets during their formation, which would then lower the C/O ratio formed in these calculations. The sublimation of ices will occur under preferential conditions that can be met if the final planet is located relatively close to the star (a few to 5-6 AU in this work). In this case, the thick atmosphere of H$_2$ and He of giant planets could shield the other molecules from observations.\\
			
			The values of the C/O ratio cannot be easily determined by direct observations of planetary atmospheres. Figure \ref{CO_20_plan} d and f show that a full range of C/O ratios can be achieved in the atmosphere of newborn planets with mass ratios ranging from 0 to 1, where the value zero corresponds to planets formed in the outer regions of the planetary system or rocky planets that did not accrete any gas. As suggested by \cite{Oberg2011}, this whole range of C/O ratios can be explained by the fact that volatile species are not condensing at the same location, resulting in depletion in the gas phase once they condense on the surface of grains in the disc, as shown in Fig. \ref{CO_ratio_20_nirr}. Irradiation plays a role in determining the C/O ratio in the envelope of low-mass planets located at $\sim$1 AU. Modelling radiative effects increases the C/O ratio to $\sim$0.9. Without these effects, such an increase is seen only for a few planets.
			
			 The C/O ratio in planetary envelopes is also sensitive to the time of accretion. For example, giant planets start runaway accretion after a few million years, when the only species left in the gas disc are H$_2$ and He. Thus the fraction observed for these species is high. This does not mean that these planets cannot reach a high C/O ratio. During core formation, and before the runaway accretion of gas, cores of giant planets can indeed still bind some gas, which results in accretion of gas consisting partly of C- and O-bearing molecules and can result in a high C/O ratio in the envelope. In a similar way, it is possible for terrestrial planets to be unable to bind enough gas enriched in C and O, because their core was not massive enough when these species were still in the gas phase. The C/O ratio in such planets remains very low. \\
		
		\section{Discussions}	 
			\subsection{Effect of planetary migration}
			Previous studies \citep{Oberg2011,Ali-Dib2014} computed the increase in the C/O ratio of planets formed insitu, which can lead to an overestimation of the ratio in the gas accreted by planets. The effect of planetary migration on the bulk composition of planets has a non-negligible effect ,as shown in T14 (and references therein), because it helps forming different types of planets that differ in mass, semi-major axis ranges, and composition. Since this migration affects the solid composition, we investigate the possible effect that the migration has on gas composition. For this purpose, the model is kept the same but the planetary composition is computed assuming the simulated planets are formed, remain, and accrete only at their final position in an evolving gas disc. \\
			
\begin{figure}
				\includegraphics[width=\columnwidth]{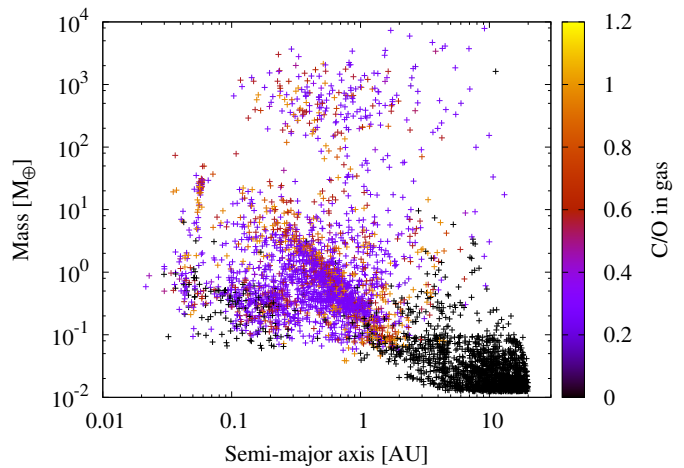}
				\caption{\label{CO_insitu_evol} C/O mass ratio in the gas accreted by planets in an \textit{in situ} scenario. The plot is shown for a model with CO/H$_2$O = 0.2, without irradiation.}
			\end{figure}

			Figure \ref{CO_insitu_evol} shows the C/O ratio obtained in such a model, without irradiation and with CO/H$_2$O = 0.2. The difference with the migration model (see Fig. \ref{CO_20_plan}.h) is rather small, except for planets located between 0.4 and 1 AU with a mass of between 0.1 and 10 M$_{\oplus}$. The migration consequently has no effect on giant (and icy) planets, which is to be expected since these planets are accreting gas mostly when (and where, for icy planets) only He and H$_2$ are in the gas phase. However, for Earth-like planets or super-Earths, the inclusion of migration reduces the C/O ratio as a result of a loss of C in their atmosphere. This depletion may play a role in the possible presence of Earth-like atmospheres for planets, if the ices in the bulk of the planets are not efficiently desorbed. 
			
			\subsection{Effect of the gas phase evolution}
				\cite{Oberg2011} showed that an increase in the C/O ratio compared to the stellar value can be explained by the fact that the different ices do not condense at the same location. However, they did not take into account the abundance evolution of volatile species in the gas phase. \cite{Ali-Dib2014} demonstrated that the C/O ratio can be increased with time, as a result of loss of species such as water, as seen in Fig. \ref{CO_ratio_20_nirr}  (see also their Fig. 4). Consequently, the C/O ratio enhancement particularly for close-in planets, might be explained by both the presence of different ice lines and the changes in abundances of the different gas species with time, although these are two different processes. \\
				
\begin{figure}
				\includegraphics[width=\columnwidth]{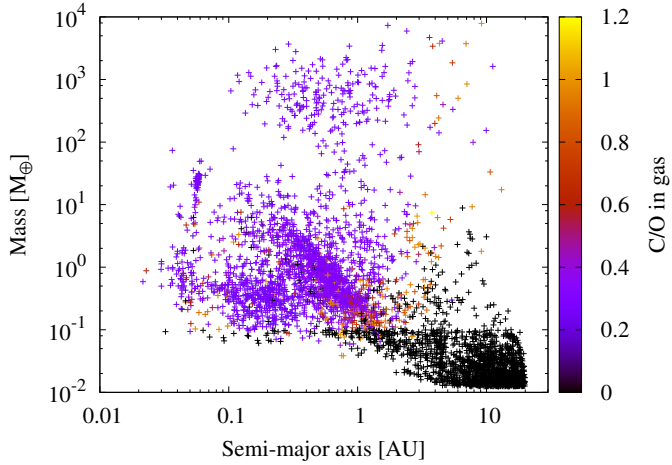}
				\caption{\label{CO_mig_noevol} C/O mass ratio in the gas accreted by planets if the gas phase of the gas does not evolve with time. The plot is shown for a model with CO/H$_2$O = 0.2, without irradiation of the disc.}
			\end{figure}

				To study the effect of such a change, we reproduced the model of \cite{Oberg2011} in which the abundance profiles remain unchanged with time. Figure \ref{CO_mig_noevol} shows the results of this simulation, with CO/H$_2$O = 0.2 and without irradiation of the disc. 
			The C/O in the gas phase is reduced for most of the planets, most of them having a ratio of about 0.2-0.3 that can be explained by the preservation of water in the disc, thus lowering the C/O ratio in the disc for planets that migrated towards the inner parts of the disc ($<$ 1AU). However, some gas giants have an increased C/O ratio for the same reasons, since none of the volatile molecules disappear from the gas phase. As discussed in \cite{Oberg2011}, we find that it is possible to explain the differences in the C/O ratios compared to the solar value only by taking into account the differences in condensation, although the diversity in the planetary ratios is weaker than in the model using evolution of the disc and planetary migration. Taking into account the evolution of the gas phase is important for giant planets as well as for planets of lower masses ($<$50 M$_{\oplus}$) located within 1-2 AU from their central star.	\\	
				A similar study by \cite{Madhusudhan2014} attempted to link the observed C/O ratio in hot Jupiters with the formation of such planets. In their model, the gas phase is similar to that obtained by \cite{Oberg2011} and does not evolve with time. \cite{Madhusudhan2014} studied three cases of migration and formation of planets: The planet formed by core accretion and migrated through (a) disc migration, (b) disc-free migration, and planets are formed (c) by gravitational instability. In this case, the results obtained by \cite{Madhusudhan2014} in their case (a), which correspond best to the simulations presented here, are very similar to what is obtained with our model without gas evolution (see Fig. \ref{CO_mig_noevol}). It is indeed difficult in the present work to form hot Jupiters with a C/O ratio that is supersolar if the gas phase evolution is switched off. In all the Jupiters within 1 AU formed in the model of A13, the C/O ratio is subsolar around 0.2-0.3. However, if the evolution of the gas phase is taken into account, these results are not true anymore, and it becomes possible to enhance the C/O ratio, which then does not constrain the formation process, as suggested by \cite{Madhusudhan2014}.			
			
			Note that in this and the previous section, planets can be formed with a high C/O ratio at semi-major axes lower than 0.2 AU. This result is different from \cite{Oberg2011}. It is due to planetary migration (this section), the gas phase evolution (previous section), and the presence of different discs.
			
			\subsection{Planet formation with different C/O ratios in the same system}
			Observations of the stellar system HR 8799 \citep{Barman2011,Konopacky2013} have shown the presence of several gas giants, two of which could have a rather different C/O ratio in the atmosphere. HR 8799c has a C/O ratio estimated to be 0.65 \citep{Konopacky2013}, and the C/O ratio in the atmosphere of HR 8799b could differ significantly from HR 8799c because of its location. Forming two gas giants in the same system but with different C/O ratios is explained by \cite{Oberg2011}, who suggested that such planets are formed \textit{in situ} in two different regions. In some of the simulated systems, we are able to obtain a system formed of at least two giant planets with different C/O ratios. Note, however, that the star considered in this work (solar mass and luminosity) and H R8799 are different and that the distances are very different as well.\\			
			In the simulated system, planet A has a final mass of 845.5 M$_{\oplus}$ and a final position of 1.23 AU and a C/O ratio in the envelope of 0.75. Its initial position is 4.3 AU. Planet B has a final mass of 570.1 M$_\oplus$ with a final position of 0.29 AU starting at 3.93 AU and a C/O ratio in the envelope of 0.9.
			
\begin{figure}
				\includegraphics[width=\columnwidth]{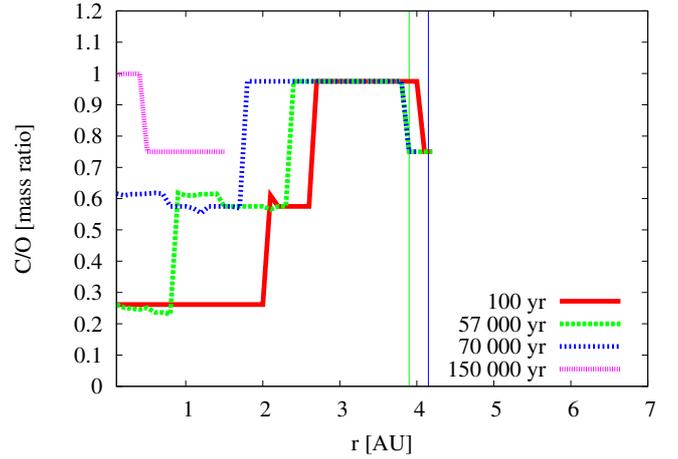}
				\caption{\label{237_COdisc} C/O mass ratio in the gas disc of the system in which planets A and B formed for different times (100, 57 000, 70 000 and 150 000 yr).  The vertical lines show the starting position of planets A (blue) and B (green). The plot is shown for a model with CO/H$_2$O = 0.2, without irradiation of the disc.}
			\end{figure}

			Figure \ref{237_COdisc} shows the evolution of the C/O ratio in the gas disc in which the two planets formed. The initial difference can be explained by the condensation of species as shown in Fig. \ref{237_COdisc}. Planet B starts at a location where CH$_4$ and CO are the only C- and O-bearing species left in the gas phase, while planet A starts with a gas consisting of CO (CH$_4$ condenses at 4.1 AU). \\
			
\begin{figure}
				\includegraphics[width=\columnwidth]{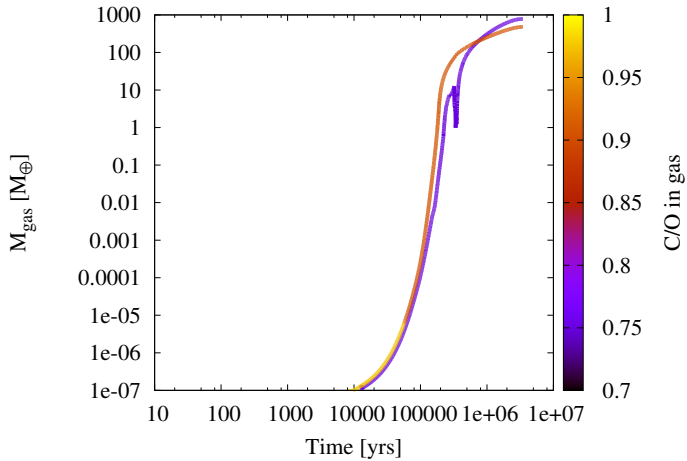}
				\caption{\label{237_evol} Evolution of the C/O ratio in the envelope as a function of time and of the mass of the envelope for planets A and B. The C/O ratio of the envelope of planet A stays constant, while the C/O ratio of planet B decreases between 60 000 and 100 000 yr, as described in the text.}
			\end{figure}

\begin{figure}
				\includegraphics[width=\columnwidth]{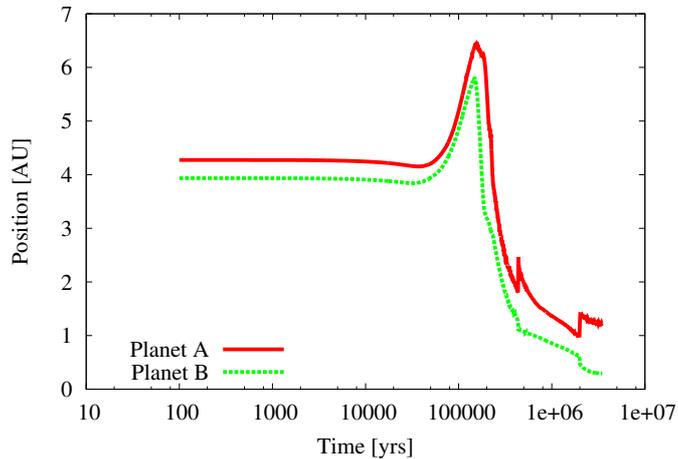}
				\caption{\label{237_migration} Evolution of the positions of the planets A and B through time.}
			\end{figure}

			Figure \ref{237_evol} shows the evolution of the C/O ratio in the envelope accreted by planets A and B as a function of time and of the mass of the envelope. Before 60 000 years, planet B evolves between 3.9 and 4 AU, where and when the C/O is as high as 1 in the gas phase. Between t=60 000 yr and t=100 000 yr, the planet migrates towards higher semi-major axes (see Fig. \ref{237_migration}), where CH$_4$ condenses, which lowers the C/O ratio to 0.9. Between t=100 000 yr and t=150 000 yr, the planets migrates to 6.5 AU, where CO condenses as well, H$_2$ and He being the only two species left in gas phase, so that the C/O ratio stays constant. Finally, after 150 000 years, the gas has evolved enough and consists only of H$_2$ and He (all other molecules have been accreted onto the star), and the C/O ratio does not change anymore. \\
			
			The initial difference in the C/O ratio of the accreted gas is thus directly linked to the location of the protoplanet when it starts accreting gas and can be explained by the presence of different ice lines for the different volatile species, as described in \cite{Oberg2011}. However, this initial ratio can decrease or increase during the migration of the protoplanet, as shown in this section for planet B at 60 000 yr; or stay roughly constant because of the evolution of the gas phase (both planets after 150 000 yr). 
			
			\subsection{Replenishment by evaporation of icy mantles on dust grain}
			Although for simplicity the planetesimals considered in this work are assumed to be of kilometer size so they do not drift, in reality, some smaller dust grains are present and drift. They will reach a region of the disc in which the temperature is high enough to evaporate their icy mantles.
			
			This process has been included in the model of \cite{Ali-Dib2014}. They found that the vapor concentration is only enhanced in a very narrow region for H$_2$O and CO. Moreover, they found that the diffusion of the gas phase is much faster than the replenishment, which implies that the latter is not high enough to counter the diffusion. The long coagulation timescales involved in the process deplete the gas amount inside the iceline. The quasi-depletion state for H$_2$O (respectively CO) in \cite{Ali-Dib2014}  is reached after $\num{2e4}$ years (respectively $\num{1e5}$ years), times that are consistent, yet much shorter, with our work. These results have also been obtained in other studies with different parameters for the disc \citep[see, e.g.][- and references therein]{Cyr1998,Ciesla2006,Ciesla2009}, but with a value of $\alpha$ similar to our simulations ($\sim$ 10$^{-3}$). 
			
			The impact of this effect is therefore probably very weak in the present calculations.
					
	\section{Summary and conclusions}	
		We presented results from extending the model of T14 and M14a,b following the approach of \cite{Oberg2011} and \cite{Ali-Dib2014} by adding the effect on the gas composition of ice lines for different gases within an evolving gas disc. Accretion of gas consisting of H$_2$, He, and eight main volatile molecules (H$_2$O, CO, CO$_2$, NH$_3$, N$_2$, CH$_3$OH, CH$_4$, and H$_2$S) was added to the model of M14a,b and T14, and we studied the physical effects, that is, gas evolution, migration of planets, and presence of ice lines, that might explain the differences in C/O between the star and observed planets. \\
		
		The simulation results are similar to the results presented by \cite{Oberg2011}. A key observation is that the different positions of ice lines enable the gas to develop a higher C/O ratio than that of the host star. The evolution of the gas phase as suggested by \cite{Lynden-Bell1974} shows that the C/O ratio in the disc also evolves with time, depending on the species present in the disc. For example, the removal of H$_2$O from the gas phase increases the C/O ratio in the gas phase of the disc to 2-3 times the stellar value. We showed that the resulting C/O ratio in planetary envelopes just after their formation is affected by
		\begin{itemize}
			\item The position of the protoplanet when it starts accreting gas, which gives the planet its initial C/O.
			\item The gas phase evolution, which depletes or enhances specific regions in volatile compounds and thus increases or decreases the initial C/O ratio. If the gas phase does not evolve \citep[as in][]{Oberg2011}, only a few planets are able to obtain a C/O ratio higher than that of the host star. 
			\item The time of accretion. The time evolution of the gas phase shows that removing species from the gas phase such as H$_2$O can increase the C/O ratio in the gas phase of the disc in the inner regions.
			\item The migration path of the protoplanet. The C/O ratio in \textit{in situ} formed planets is slightly different for rocky and giant planets, but the effect is strong for planets located between 0.4 and 1 AU whith a mass of between 0.1 and 10 M$_{\oplus}$. However, the migration path of the protoplanet can increase or decrease its initial C/O ratio, depending on the gas phase evolution.
			\item The irradiation. Irradiation pushes the position of the ice line outwards (see T14, M14a), meaning that the volatile molecules can be present at positions farther outside than in evolving planetary systems without irradiation. This leads to a possible enrichment of the C/O ratio in the envelope of low-mass planets around 1 AU.
		\end{itemize}
		We also showed that if ices are assumed to fully sublimate into the gas phase, the effect of the accreted gas on the C/O ratio is negligible compared to the contribution of the solids. This suggests that high values of the C/O ratios, as observed in giant planets, are due to a lack of exchange of material between the planet core and its envelope (sublimation of ices or condensation). Another explanation could be that the observations are biased because the envelope of a planet is probably composed of several layers and that observations are not taking into account this layered-composition. Finally, the results show that the observed C/O ratio in atmospheres may not be representative of the total C/O ratio of the planet.
		
		\paragraph{{Acknowledgments}} This work was supported by the European Research Council under grant 239605, the Swiss National Science Foundation, and the Center for Space and Habitability of the University of Bern. This work has in part been carried out within the frame of the National Centre for Competence in Research PlanetS supported by the Swiss National Science Foundation.

	\bibliographystyle{aa}
	\bibliography{biblio}

\end{document}